\documentclass[10pt,journal,compsoc]{IEEEtran}

\ifCLASSOPTIONcompsoc
  \usepackage[nocompress]{cite}
\else
  \usepackage{cite}
\fi

\usepackage{cite}
\usepackage{amsmath,amssymb,amsfonts}
\usepackage{graphicx}
\usepackage{textcomp}
\usepackage{xcolor}
\usepackage[hyphens]{url}
\usepackage{array}
\usepackage[hidelinks]{hyperref}
\usepackage{dblfloatfix}
\usepackage{placeins}
\usepackage{siunitx}
\usepackage{caption}
\usepackage{subcaption}
\usepackage{enumitem}
\usepackage{hhline}
\usepackage{tikz}
\newcommand*\circled[1]{\tikz[baseline=(char.base)]{
            \node[shape=circle,draw,inner sep=0.8pt] (char) {#1};}}

\usepackage[colorinlistoftodos]{todonotes}
\usepackage{fancyhdr}
\usepackage[normalem]{ulem}
\usepackage{xspace}
\usepackage{tabularx}
\usepackage{booktabs} 
\usepackage{multirow}
\usepackage{pifont}
\newcommand{\cmark}{\ding{51}}%
\newcommand{\xmark}{\ding{55}}%

\begin{document}



\newcommand{\rev}[1]{{\color{black}#1}} 
\newcommand{\revv}[1]{{\color{black}#1}} 

\newcommand{\commname}{COALA\xspace}
\newcommand{\alloname}{WACO\xspace}

%
\title{Stream: Design Space Exploration of Layer-fused DNNs on Heterogeneous Dataflow Accelerators}
%
%
%
%

\author{Arne~Symons,
        Linyan~Mei,
        Steven~Colleman,
        Pouya Houshmand,
        Sebastian~Karl,
        and~Marian~Verhelst,~\IEEEmembership{Fellow,~IEEE}
\IEEEcompsocitemizethanks{\IEEEcompsocthanksitem The authors are with MICAS, The Department of Electrical Engineering
(ESAT), KU Leuven, 3001 Leuven, Belgium.
E-mail: arne.symons@kuleuven.be.}
}

%
%

\markboth{IEEE TRANSACTIONS ON COMPUTERS,~Vol.~74, No.~1, January 2025}%
{
}
%


\IEEEtitleabstractindextext{%
\begin{abstract}
As the landscape of deep neural networks evolves, heterogeneous dataflow accelerators, in the form of multi-core architectures or chiplet-based designs, promise more flexibility and higher inference performance through scalability. So far, these systems exploit the increased parallelism by coarsely mapping a single layer at a time across cores, which incurs frequent costly off-chip memory accesses, or by pipelining batches of inputs, which falls short in meeting the demands of latency-critical applications. To alleviate these bottlenecks, this work explores a new fine-grain mapping paradigm, referred to as layer fusion, on heterogeneous dataflow accelerators through a novel design space exploration framework called \emph{Stream}.

\emph{Stream} captures a wide variety of heterogeneous dataflow architectures and mapping granularities, and implements a memory and communication-aware latency and energy analysis validated with three distinct state-of-the-art hardware implementations. As such, it facilitates a holistic exploration of architecture and mapping, by strategically allocating the workload through constraint optimization. The findings demonstrate that the integration of layer fusion with heterogeneous dataflow accelerators yields up to $2.2 \times$ lower energy-delay product in inference efficiency, addressing both energy consumption and latency concerns. 

The framework is available open-source at: \href{https://github.com/kuleuven-micas/stream}{github.com/kuleuven-micas/stream}.
\end{abstract}

}

\maketitle

\IEEEdisplaynontitleabstractindextext

%
\IEEEpeerreviewmaketitle

\ifCLASSOPTIONcompsoc
\IEEEraisesectionheading{\section{Introduction}\label{sec:introduction}}
\else
\section{Introduction}
\label{sec:introduction}
\fi

\IEEEPARstart{D}{eep} neural networks (DNNs) have significantly advanced various fields, including computer vision, natural language processing, and signal analysis. However, this progress has come with a substantial increase in the size of networks (weights) and intermediate data values (activations). Simultaneously, the demand for edge processing has risen, imposing strict energy and latency requirements with a constraint on memory footprint.

\subsection{\rev{Dataflow Accelerator Architectures}}
\begin{figure}
\centering
\includegraphics[width=\columnwidth]{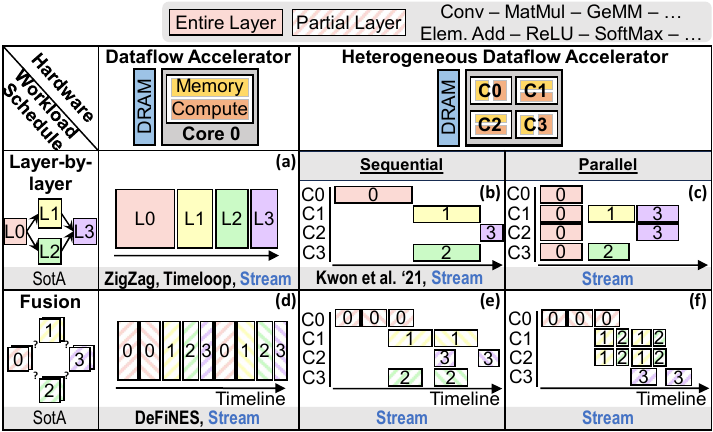}
\caption{\rev{A conceptual example showing different ways of scheduling a deep neural network workload onto different hardware accelerators.}}
\vspace{-1em}
\label{fig:general_concept}
\end{figure}

Dataflow accelerator (DA) architectures have evolved from single-core designs with specific dataflows~\cite{ShiDianNao,Eyeriss,Envision} to multi-core and chiplet-based systems~\cite{Planaria, Simba, Illusion, verma_jssc, shin2018dnpu}, increasing computing parallelism. Heterogeneous dataflow accelerator (HDA) architectures offer specialized processing for diverse DNN layer types~\cite{Kwon2021, diana, garofalo2022heterogeneous}, employing varied dataflows, PE array sizes, and memory capacities. Fig.~\ref{fig:general_concept} illustrates DNN inference scheduling on DA and quad-core HDA architectures. Fig.~\ref{fig:general_concept}(a-c) shows \emph{layer-by-layer} or \emph{layer-first} scheduling, which can improve throughput through pipelining but fails to enhance latency due to under-utilized cores (Fig.~\ref{fig:general_concept}(b)).

\subsection{\rev{Fine-Grained Scheduling Strategies}}
Parallelizing a layer across multiple cores for increased core utilization introduces challenges for HDA architectures, as shown in Fig.~\ref{fig:general_concept}(c). An alternative approach is \emph{layer-fused} scheduling~\cite{LayerFusion}, also known as \emph{depth-first}~\cite{DepthFirst} or \emph{cascaded}~\cite{TVM_cascade} scheduling (Fig.~\ref{fig:general_concept}(d-f)). This technique processes a segment of outputs sequentially through a fused stack of layers, reducing memory footprint and off-chip data accesses while enhancing workload parallelism (Fig.~\ref{fig:general_concept}(e)). Combined with layer parallelization (Fig.~\ref{fig:general_concept}(f)), this improves core utilization but increases core-to-core transfers. Existing layer-fused scheduling techniques are often hardware-specific~\cite{depfin, diana, liu2020dlfusion, colleman2021high}, limiting performance evaluation across diverse scenarios. The mapping of layer-fused DNNs onto various HDA architectures remains largely unexplored.

\subsection{Contributions of this Work}
This work introduces the first comprehensive exploration framework for HDA architectures, featuring fine-grained scheduling of layer-fused DNNs. The primary contributions are as follows:

\begin{itemize}
    \item \textbf{Stream}: An open-source design space exploration framework called \emph{Stream} is presented for the mapping of layer-fused DNNs on heterogeneous dataflow accelerator architectures. Stream offers a flexible mapping granularity and architecture representation, facilitating adaptable and comprehensive exploration across diverse dataflow strategies and architectural configurations through a rapid fine-grain data dependency extraction (Section~\ref{sec:implementation}).
    \item \textbf{\commname}: The hardware performance of fine-grain mappings onto these architectures is estimated through a memory- and communication-aware latency analysis (\emph{\commname}). The accuracy of \commname is validated through comparison with three state-of-the-art implementations of hardware accelerators architectures that employ layer fusion (Section~\ref{sec:scheduler}).
    \item \textbf{\alloname}: The allocation and scheduling of the fine-grain workload is optimized through constraint optimization (\emph{\alloname}). \alloname achieves up to $32 \%$ better Energy-Delay Product (EDP) compared to a SotA genetic algorithm-based allocator due to its ability to partition layer parts across cores. Further experiments reveal up to a $2.2 \times$ EDP reduction compared to traditional layer-by-layer scheduling and show Stream's ability to evaluate a wide range of HDA architectures (Section~\ref{sec:allocation}~\&~\ref{sec:exploration}).
    
\end{itemize}

\section{Background \& Related Works}

In this section, we first provide the background on DNN acceleration using dataflow accelerator architectures. Next, we detail mapping techniques of DNNs onto these architectures. The state-of-the-art comparison is summarized in Table~\ref{tab:sota}.

\setlength\extrarowheight{2pt}
\begin{table}[ht]
\centering
\caption{Comparison of State-of-the-art frameworks.}
\label{tab:sota}
\begin{tabularx}{\columnwidth}{lcccc}
\toprule\
\textbf{Framework} & \textbf{\begin{tabular}{@{}c@{}}HDA \\ Model\end{tabular}} & \textbf{\begin{tabular}{@{}c@{}}Layer \\ Fusion\end{tabular}} & \textbf{\begin{tabular}{@{}c@{}}Optimization \\ Method\end{tabular}} & \textbf{\begin{tabular}{@{}c@{}}Perf. \\ Evaluation\end{tabular}} \\
\midrule
Kwon et al.~\cite{Kwon2021} & \cmark & \xmark & \begin{tabular}{@{}c@{}}Utilization \\ Heuristic\end{tabular} & \begin{tabular}{@{}c@{}}Analytical \\ Model\end{tabular} \\
\hline
TVM~\cite{TVM_cascade} & \xmark & \cmark & \begin{tabular}{@{}c@{}}Memory \\ Heuristic\end{tabular} & \begin{tabular}{@{}c@{}}ARM \\ Ethos-U\end{tabular} \\
\hline
DeFiNES~\cite{DeFiNES} & \xmark & \cmark & \begin{tabular}{@{}c@{}}Exhaustive \\ Tile Search\end{tabular} & \begin{tabular}{@{}c@{}}Analytical \\ Model\end{tabular} \\
\hline
DNNVM~\cite{DNNVM} & \xmark & \cmark & \begin{tabular}{@{}c@{}}Subgraph\\ Isomorphism\end{tabular} & FPGA \\
\hline
Olympus~\cite{Olympus} & \xmark & \cmark &\begin{tabular}{@{}c@{}}X-partition, \\ Dynamic Progr.\end{tabular} & Interstellar \\
\hline
TileFlow~\cite{zheng2023tileflow} & \xmark & \cmark & \begin{tabular}{@{}c@{}}Tree-based \\ Analysis\end{tabular} & \begin{tabular}{@{}c@{}}Analytical \\ Model\end{tabular} \\
\hline
\begin{tabular}{@{}l@{}}This Work \\ (Stream)\end{tabular} & \cmark & \cmark & \begin{tabular}{@{}c@{}}Constraint \\ Optimization\end{tabular} & \begin{tabular}{@{}c@{}}Analytical \\ Model\end{tabular} \\
\bottomrule
\end{tabularx}
\vspace{-1em}
\end{table}

\subsection{\rev{Dataflow Accelerator Architectures}}

\begin{figure}
\centering
\includegraphics[width=\columnwidth]{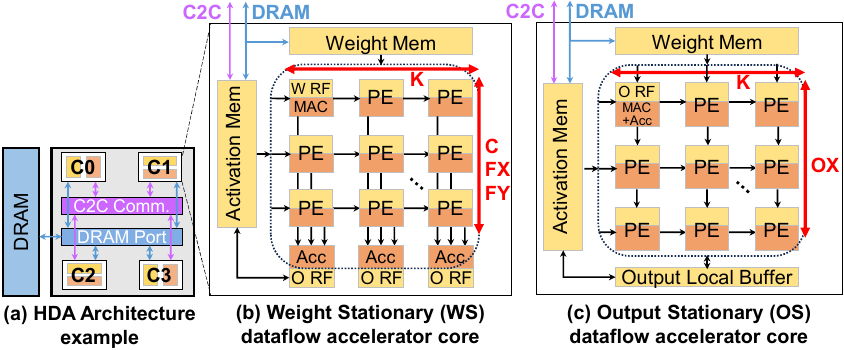}
\caption{Architectural schematics of dataflow accelerators: (a) Heterogeneous Dataflow Accelerator (HDA), (b) Weight Stationary (WS) core, and (c) Output Stationary (OS) core.}
\label{fig:multi_core_architecture}
\vspace{-1.5em}
\end{figure}

Traditional accelerator cores, as shown in Fig.~\ref{fig:multi_core_architecture}(b), utilize arrays of spatially unrolled processing elements (PEs) to enable parallel computations. Depending on the dataflow, different tiles of the DNN are paralellized on the PE array. To further increase compute parallelism while maintaining efficiency, the field has evolved towards multi-core or multi-chiplet-based architectures (Fig.~\ref{fig:multi_core_architecture}(a)). This trend began with homogeneous multi-core systems, which replicate the same accelerator across multiple units. Examples include Planaria~\cite{Planaria} with its partitionable systolic array, Illusion~\cite{Illusion}'s 8-chip design with minimal on-chip memories, Simba~\cite{Simba}'s 36-chiplet architecture, and the prototype of a $4 \times 4$ clustered analog in-memory compute (AiMC) system~\cite{verma_jssc}.

Building on this foundation, heterogeneous dataflow accelerators (HDA) designs, such as those explored in~\cite{Kwon2021} and implemented in DIANA~\cite{diana}, incorporate multiple sub-accelerators (i.e. cores) with varying dataflow specializations. By combining different core types (like digital and AiMC cores) and dataflows, HDAs offer more tailored and efficient processing for a wider range of DNN layer types, addressing the growing diversity in neural network architectures.

\subsection{Workload Allocation, Scheduling \& Mapping}
\label{background_mapping}

The recent trend towards multi-core accelerator systems enlarges the traditional mapping space. It is divided into three categories:

\subsubsection{Allocation} 
Each layer of the workload must firstly be allocated to one or more cores (from now on referred to as \textit{workload allocation}). Allocating a modern DNN of 50 layers onto e.g. a quad-core architecture yields $10^{30}$ possible workload allocations. Kwon et al.~\cite{Kwon2021} allocate the workload to the heterogeneous cores through a utilization-based heuristic, \rev{but their work assumes a fixed communication topology through a shared buffer, which can incur unnecessary energy and latency overhead for small tensors.}

\subsubsection{Scheduling} 
Each allocated layer must subsequently be scheduled in time. This determines the execution order of the layers (or their fine-grained tiles). In traditional layer-by-layer processing, the only scheduling flexibility comes from the possibility to reorder branches in the DNN. More scheduling opportunities, however, arise when scheduling at a sub-layer granularity. Instead of processing an entire layer at once, a smaller tile of each layer is processed and its outputs are immediately consumed to process tiles of the subsequent layers. The size of such a layer tile determines the \emph{scheduling granularity}. 

The scheduling of layer-fused DNNs presents its own set of challenges and opportunities. Challenges include diminished data reuse opportunities as layer tiles become smaller, together with a large increase in the number of valid scheduling orderings due to the diversity of the workload allocation on the HDA architectures. Despite these challenges, a careful mapping of layer-fused DNNs on HDA architectures can drastically improve the available workload parallelism and the retention of more data on-chip, thanks to the reduced data requirements per layer tile. 

TVM~\cite{TVM_cascade} includes a cascading scheduler that explores different scheduling granularities, but is tied to the ARM Ethos-U microcontroller platform. DeFiNES~\cite{DeFiNES} introduces an analytical cost model to enable fast depth-first scheduling design space exploration for single-core DA architectures. Yet, the exploration of layer fusion on HDA architectures remains largely unexplored. DNNVM~\cite{DNNVM} optimizes the scheduling through a number of heterogeneous compiler optimizations for FPGA-accelerated inference. Olympus~\cite{Olympus} reaches memory-optimality for a single-core FPGA-based accelerator through layer fusion. In TileFlow~\cite{zheng2023tileflow}, a tree-based representation of nested for loops achieves optimal layer fusion for a homogeneous multi-core accelerator architecture using shared memories.

\subsubsection{Mapping}
Lastly, the computations of each layer tile must be efficiently mapped onto each core, respecting the constraints stemming from the supported dataflow and memory resources of each individual core.
The spatial dataflow plays a key role in the core's utilization, as a mismatch between the core's dataflow dimensions and the tile granularity will cause spatial under-utilization of the PE array. The temporal mapping, on the other hand, impacts the temporal data reuse from the different memories of the core's memory hierarchy~\cite{verhelst2017embedded, LOMA}. 
Many design space exploration (DSE) frameworks have arisen~\cite{Interstellar, MAESTRO, Timeloop, ZigZag, CoSA, MindMap, GAMMA} to analytically estimate the HW cost and optimize the efficiency of the spatial and temporal mapping through loop optimizations such as unrolling, tilling and reordering. These frameworks model a layer-by-layer scheduling, summing up the energy and latency cost across layers.


\section{Stream framework}
\label{sec:implementation}

\section*{Motivation \& Overview}

The increasing complexity of HDA architectures, characterized by the number of cores, each core's dataflow, memory distribution, and inter-core connection topology and bandwidth, significantly impacts hardware performance.
Concurrently, the scheduling of layer-fused DNNs presents its own aforementioned challenges and opportunities.

To date, no framework exists that facilitates this co-exploration, hindered by the difficulty in rapidly extracting the fine-grain data dependencies between layer-fused tiles, accurately modeling constraints and overheads from the multi-core hardware architectures, and efficiently exploring the vast workload allocation space. Moreover, it is crucial that the framework is easily extendable to a wide variety of future architectures and scheduling paradigms. This motivates the need for a novel design space exploration framework: \emph{Stream}. Stream aims to leverage the parallelization opportunities and address the challenges posed by both HDA architectures and fine-grain layer fusion, with accurate hardware performance modelling.

\begin{figure}[!t]
\centering
\includegraphics[width=\columnwidth]{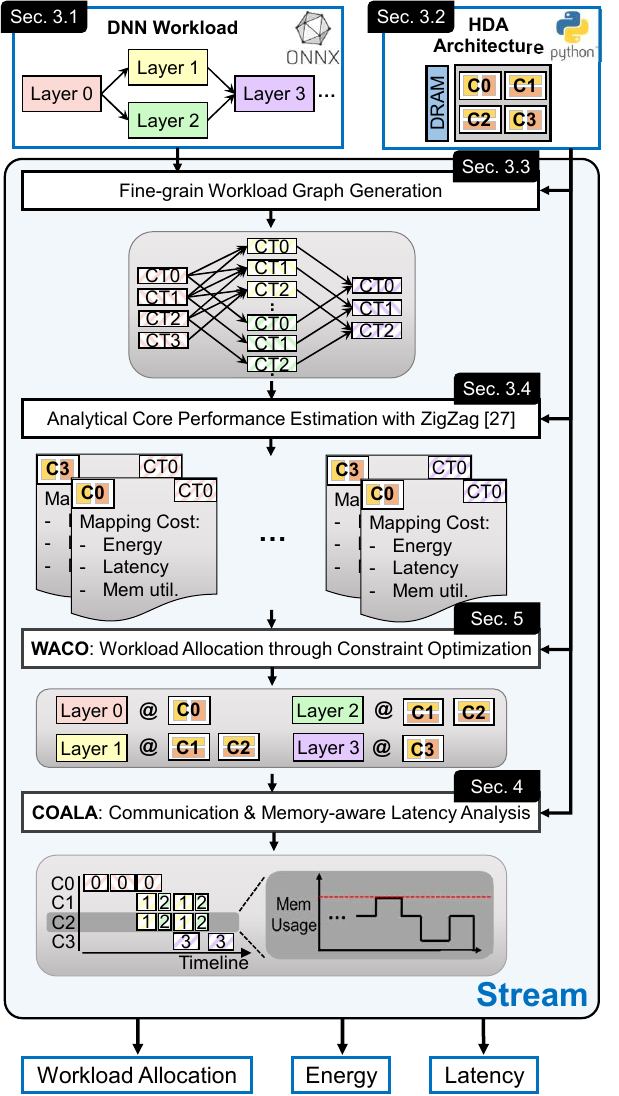}
\caption{Overview of the Stream framework.}
\vspace{-1em}
\label{fig:overview}
\end{figure}

Fig.~\ref{fig:overview} shows an overview of Stream. The DNN workload is first converted into a fine-grain workload graph of layer tiles based on the desired granularity and core dataflows. Next, the performance of each unique tile-core combination is estimated using an analytical single-core DSE framework, cfr. Section~\ref{background_mapping}. 
Given a workload allocation, which is optimized in the \alloname engine, \commname estimates the latency and energy consumption of the DNN inference onto the HDA architecture, taking into account data communication and limited on-chip memory. 

\section*{Implementation}



\subsection{Input: ONNX Workload}

Stream converts each operator of the provided ONNX~\cite{ONNX} model into a layer containing a set of nested for-loops of different dimensions and loop bounds. Each layer can contain a stride in any dimension and as a consequence Stream is capable of representing all popular DNN operators: Conv1d, Conv2d, Conv3d, MatMul, GeMM, Pooling, Softmax, Add, etc. The layers are interconnected in a directed acyclic graph. 

\subsection{Input: HDA Architecture}

The HDA architecture definition within Stream is also designed with versatility and adaptability at its core. \revv{A core is characterized by a PE array shape and a hierarchical memory structure. Each memory is defined with a specified capacity, a designated number of read, write, or read/write ports, and read/write energy costs. The ports are constrained by a limited bandwidth to simulate realistic hardware limitations accurately.}

\begin{figure}[!t]
\centering
\includegraphics[width=1\columnwidth]{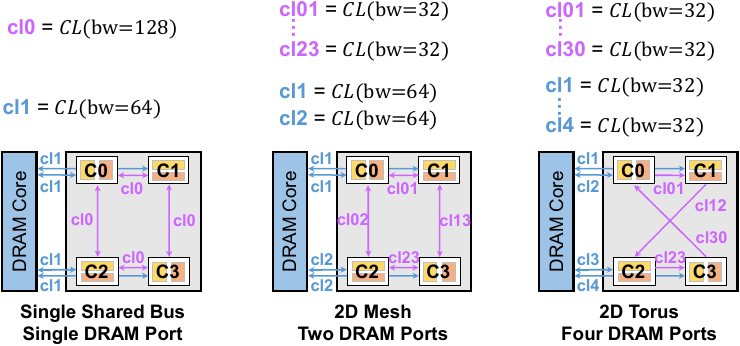}
\caption{Stream models a graph of cores, with CommunicationLink objects attached to its edges, to represent a wide range of HDA architectures. }
\vspace{-1em}
\label{fig:hda_model}
\end{figure}

\rev{The cores in the HDA are interconnected in a directed graph \(G = (V, E)\), where \(V\) is the set of vertices representing the cores, and \(E\) is the set of directed edges representing the communication links between these cores. Each edge \(e \in E\) has an associated Communication Link (CL), which embodies the communication channel between cores.} These CLs are characterized by a limited bandwidth, which mirrors the communication constraints found in actual hardware designs. This graph-based representation of core interconnectivity allows the HDA architecture to effectively emulate a wide range of modern core-to-core and off-chip communication topologies. Fig.~\ref{fig:hda_model} illustrates the flexibility using three examples ranging from a simple single shared bus architecture to sophisticated 2D mesh and torus configurations with varying off-chip communication ports.

\subsection{Fine-grain Workload Graph Generation}
\label{sec:attribute_extraction}

\rev{\subsubsection{Layer Fusion Approach}

Layer fusion in DNN models can be implemented through different strategies, such as integrating a primary operation with an auxiliary element-wise function (e.g., combining a convolutional layer with ReLU) or merging multiple core operations (e.g., consecutive convolutional layers). The Stream framework combines both, as its goal is to explore the possibilities of deep fusion, where tiles of layer activations are reused in a depth-first manner through the workload. This is achieved through the combination of partitioning each layer of the workload into finer computation tiles (\revv{CT}) and inferring the dependencies between them. Moreover, the \alloname engine (Section~\ref{sec:allocation}) smartly incorporates the hardware constraints by limiting the fusion depth to not exceed the weight memory capacity of the HDA cores.
}

\begin{figure}[!t]
\centering
\includegraphics[width=0.9\columnwidth]{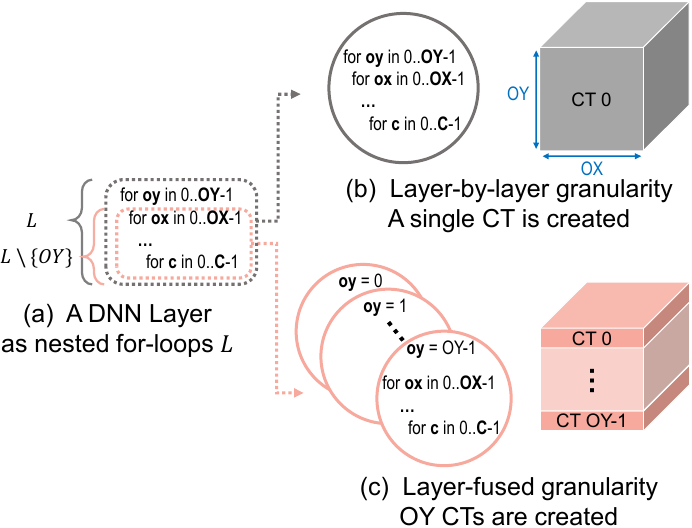}
\caption{\rev{Each layer is partitioned into one or more Computation Tiles (\revv{CT}) depending on the desired granularity.}}
\vspace{-1em}
\label{fig:CT_granuarity_impact}
\end{figure}

\subsubsection{Computation Tile Granularity}
The first step of representing a layer-fused DNN is splitting each layer into multiple individually schedulable tiles, referred to as \revv{computation tiles (CT)} in the remainder of this work. \rev{Assume the layer \(L\) consists of the following nested for-loops, \revv{where $l_i$ represents the loop index controlling the iteration over the dimension \( D_i \), and $L_i$ is its upper bound}}:

\[
\begin{aligned}
\text{for } l_1 &\in \text{range}(L_1): \\
&\quad \text{for } l_2 \in \text{range}(L_2): \\
&\quad \quad \vdots \\
&\quad \quad \quad \text{for } l_n \in \text{range}(L_n): \\
&\quad \quad \quad \quad O[...] \mathrel{{+}{=}} A[...] \times B[...]
\end{aligned}
\]

\revv{For example, for 2D convolutional layers, the \( D_i \) are commonly \( OX \), \( OY \) for the spatial output activation dimensions, \( FX \), \( FY \) for the spatial weight dimensions, and  \( C \), \( K \) for the input and output channels.} These for-loops can be represented as a set. A \revv{CT} is defined as a subset of \(L\):

\begin{equation*}
    L = \{l_1, l_2, \ldots, l_n\}; \quad \quad \revv{CT} \subseteq L
\end{equation*}

\revv{Note that this definition does not enforce any ordering of the loops. The ordering will be optimized for the different HDA cores.} The elements of \revv{CT} represent the loops that are nested within the tile, while the loops in \(L \setminus \revv{CT}\) \revv{(i.e., the loops in \(L\) that are not part of the computation tile)} are considered outer-\revv{CT} loops. The outer-\revv{CT} loops determine the number of \revv{CTs} generated for each layer. This is illustrated in Fig.~\ref{fig:CT_granuarity_impact}. Stream allows for manual definition or automatic inference of the \revv{CT} granularity. In traditional layer-by-layer mode, each layer is converted to a single \revv{CT} (Fig.~\ref{fig:CT_granuarity_impact}(b)). In layer-fused mode, Stream splits each layer into the number of output rows ($OY$) with optional partitioning into the output channel dimension ($K$), while taking into account that each \revv{CT} should maintain enough $K$ loops needed for the dataflows of the HDA (Fig.~\ref{fig:CT_granuarity_impact}(c)). This ensures good spatial utilization when mapping \revv{CTs} to different cores of the system. \revv{While the framework supports other layer fusion splits, they are not covered in this work.}

\subsubsection{Fine-grain Data Dependencies}
\label{sec:dependency_generation}
After the identification of the \revv{CTs} of each layer, the data dependencies between all \revv{CTs} must be generated in order to schedule them \revv{correctly and with structured memory accesses}. This process is split into two parts:

\rev{\textit{Intra-layer:} First, the intra-layer \revv{CT} dependency edges are inserted based on the outer-\revv{CT} loop order. This approach serves several purposes: it ensures that the required data accesses of \revv{CTs} within a layer are structured and hardware-friendly, as enforcing a consistent execution order avoids unstructured memory access patterns; moreover, it reduces redundant degrees of freedom in the scheduling process, streamlining the overall design while maintaining optimality, as the core already executes the \revv{CTs} of a layer sequentially; and it maintains consistent scheduling across dimensions (e.g., the batch dimension) for all layers, ensuring regularity in execution and efficiency in fusion.}

\begin{figure}[!t]
\centering
\includegraphics[width=\columnwidth]{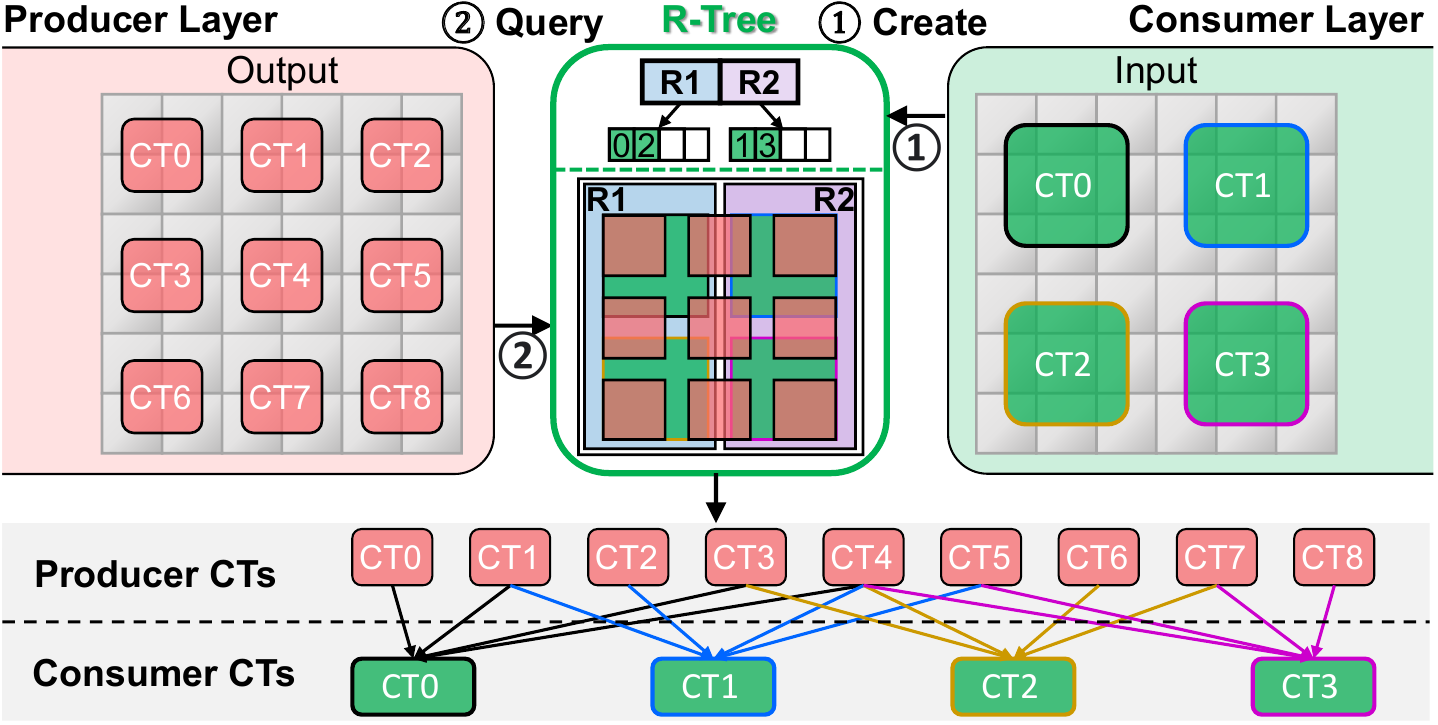}
\caption{Inter-layer \revv{CT} dependency generation example using R-trees~\cite{RTrees}.} 
\vspace{-1em}
\label{fig:CT_dependency_generation}
\end{figure}

\textit{Inter-layer}: Inter-layer \revv{CT} dependencies are determined by identifying overlaps in data generated by \revv{CTs} of one layer and consumed by the next. Given the fine-grained scheduling in modern DNNs, with up to $10^{6}$ \revv{CTs}, a direct pairwise check for dependencies would require $10^{12}$ operations, which is impractical. Instead, an efficient \textit{R-tree}~\cite{RTrees} algorithm is used to quickly identify these dependencies.
Fig.~\ref{fig:CT_dependency_generation} illustrates this process. First, an R-tree is constructed for the consumer layer’s \revv{CTs} (\circled{1}). Then, the R-tree is queried with each producer \revv{CT} to find overlapping consumer \revv{CTs} (\circled{2}). Although the example shown is simplified with 4 non-overlapping \revv{CTs}, the method supports multi-dimensional overlapping ranges.

This R-tree-based approach greatly speeds up a baseline pairwise algorithm by three orders of magnitude, reducing dependency generation time from over 9~hours to 6~seconds.

\subsection{Analytical Core Performance Estimation}
\label{sec:core_perf_est}
In this step latency performance, energy consumption and memory requirements of executing individual \revv{CTs} on each of the available HDA cores are extracted. As discussed in Section~\ref{background_mapping}, multiple DSE frameworks already exist for optimizing the mapping of layer-by-layer workloads onto single-core HW architectures~\cite{Interstellar, Timeloop, ZigZag, GAMMA, CoSA, MindMap}. These frameworks model the hardware performance of a layer and optimize the spatial dataflow and temporal mapping for the targeted accelerator, in the form of loop ordering and tiling in the memory hierarchy.

Stream exploits ZigZag~\cite{ZigZag} to obtain the optimal intra-core mapping for each \revv{CT} on each available core along with the associated hardware costs. This is achieved by parsing the for loops embedded within a \revv{CT} (Fig~\ref{fig:CT_granuarity_impact}(c)) into the layer representation of ZigZag. If not all data can be tiled on-chip, the off-chip memory instance is added to the core's hierarchy. ZigZag provides a detailed latency model for both the on- and off-loading of data to/from off-chip memory. Moreover, it models data stalls resulting from inadequate memory bandwidth within the core~\cite{mei2022latency}. The spatial dataflow and temporal mapping are optimized for latency~\cite{LOMA}.

Although this work integrates with ZigZag for core performance estimation, the flexible HDA template allows for the easy use of alternative performance estimation frameworks.

\subsection{Workload Allocation}
\label{sec:layercorealloation}

To effectively schedule the workload across time, each \revv{CT} needs to be allocated to a core. This is essential for accurately evaluating the potential parallelism and the overheads associated with off-chip and core-to-core data transfers. For now, the allocation is considered given. Section~\ref{sec:allocation} details how Stream achieves optimal workload allocation.

\section{Communication \& Memory-aware Latency Analysis with \commname}
\label{sec:scheduler}

The goal of the communication \& memory-aware latency analysis (\commname) is to estimate the latency and energy consumption of a given workload, core performances, allocation and HDA, as illustrated in Fig.~\ref{fig:coala}. Performing accurate estimations comes with two major challenges: 1) correctly assessing the core-to-core communication overhead given the HDA connectivity topology, and 2) taking into account the required amount of off-chip traffic for varying \revv{CT} granularity and on-chip memory capacity. These overheads are modelled through the help of a communication manager and a memory manager.

\subsection{Communication Manager}

Whereas modeling the communication cost of transferring activations from core to core is relatively simple for traditional layer-by-layer execution~\cite{Illusion}, the complexity grows for finer \revv{CTs} that can execute in parallel. To model transfer contention, a communication manager registers each transfer activity to \texttt{CommunicationLink} (\emph{CL}) object(s) attached to all communication channels between the cores involved in the specific data transfer (Fig.~\ref{fig:multi_core_architecture}). The transfer time $\Delta t$ depends on the packet size and link bandwidth, and the energy overhead $\Delta e$ depends on the unit energy cost. For multi-hop transfers, the communication manager finds the earliest time window in which all involved links are free and energies are accumulated.

\subsection{Memory Manager}

To know which tensor transfers are needed, a dedicated memory manager tracks the lifetime of tensors in each memory throughout the scheduling process. This allows to schedule memory eviction and reloading operations while ensuring the capacity of any of the on-chip memories is never surpassed. This work operates under the premise that data transfers always occur between the topmost memory level within each core. 

\begin{figure}[!t]
\centering
\includegraphics[width=\columnwidth]{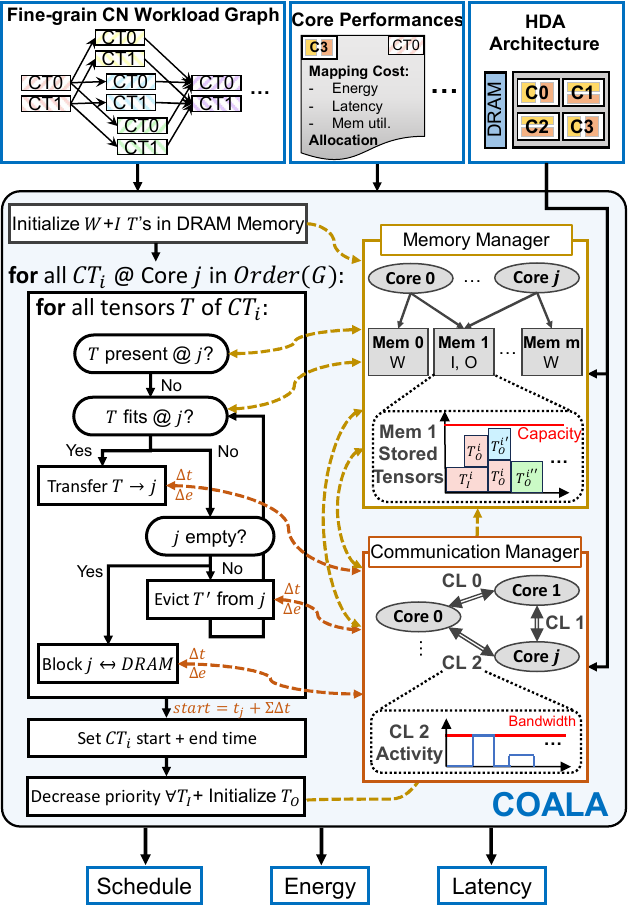}
\caption{Accurate scheduling is achieved through a communication \& memory manager that take core-to-core and off-chip transfers, and limited on-chip memory into account.} 
\label{fig:coala}
\vspace{-1em}
\end{figure}

As depicted in Fig.~\ref{fig:coala}, the decision-making process for scheduling $\revv{CT}_i$ onto its allocated core $j$ ($a_{i,j}$) as such involves several evaluations and actions by the memory manager. The \revv{CT} requires the presence of one or more input tensors $T_I$ and weight tensors $T_W$ before it can be executed. Should the tensor $T$ already reside in the on-chip memory of core $j$, no memory or communication overhead is incurred. Otherwise, the available memory space to store $T$ is assessed. In case of sufficient space, the communication manager transfers $T$, updating the appropriate CommunicationLink (CL) activities and memories. Insufficient space triggers the eviction of one or more low-priority tensors $T'$ stored in the core's memory. This priority $T'$ is stored in the memory manager and computed based on the number of unscheduled \revv{CTs} that require $T'$ and its size. This approach evicts smaller tensors first to avoid the significant impact of off-loading large tensors, aiming to optimize memory usage and scheduling efficiency. Finally, in case all tensors have been evicted and there is still insufficient space, DRAM is included as the top-level memory for the core performance estimation, and an activity is registered for the link(s) between $j$ and DRAM during the execution of $\revv{CT}_i$ for the fetching of smaller tiles of $T$.

Once all input tensors $T_I$ required for $\revv{CT}_i$ are present, the \revv{CT} is scheduled at available time $t_j$ of core $j$ plus the potential delays due to transfers $\Sigma \Delta t$. The priority of the input tensors is decreased, and the output tensor $T_O$ is initialized in the memory manager.

\subsection{Validation}
\label{sec:validation}

Stream is deployed to model the behavior of three SotA taped-out HW architectures, in order to demonstrate its modeling flexibility and to quantify its \commname latency modeling and memory footprint accuracy, compared to that of their measured system performance.

\begin{figure}
\centering
\includegraphics[width=\columnwidth]{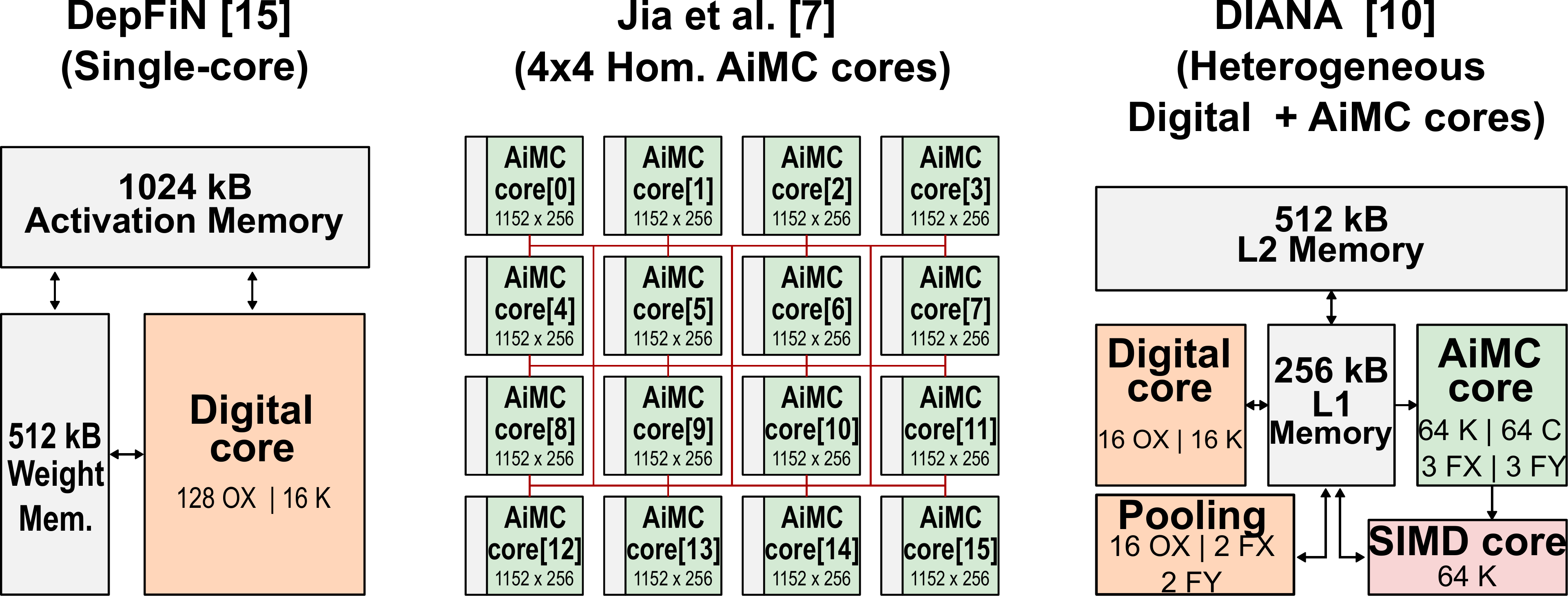}
\caption{Hardware architecture targets for the validation.}
\label{fig:validation_architectures}
\vspace{-1em}
\end{figure}

\textbf{HW Targets.} Fig.~\ref{fig:validation_architectures} shows the three architectures considered in this validation. They are diverse in their core and memory architectures and supported scheduling granularities:
\begin{enumerate}
    \item DepFiN~\cite{depfin} is a single core DNN accelerator designed for high-resolution pixel processing workloads which deploys layer fusion with line-based \revv{CTs}. The architecture includes a digital core with an $OX|K$ dataflow, \SI{512}{\kilo\byte} of weight memory and \SI{1}{\mega\byte} of activation memory. Off-chip communication is limited to $64$ bit per clock cycle.
    
    \item Jia et al.'s \cite{verma_jssc} multi-core architecture consists of a 4$\times$4 array of analog in-memory-compute (AiMC) cores, enabling high throughput and energy efficiency through pipelined execution. The cores are grouped and communicate with a $128$ bits per clock cycle on-chip network. The cores operate on a weight-stationary dataflow, with the weights loaded into the analog array before the computations. The weight loading can happen in parallel, through multiple off-chip bandwidth links of $128$ bits per clock cycle.

    \item DIANA~\cite{diana}, a heterogeneous multi-core AiMC + digital hybrid DNN accelerator SoC targets efficient end-to-end inference for edge applications. 
    The architecture includes a shared \SI{256}{\kilo\byte} L1 memory to efficiently transfer activations between cores. The digital core employs an $OX|K$ dataflow, while the AiMC core includes, similarly to Jia et al.'s cores, a weight stationary dataflow with the weights loaded into the array. The SIMD core handles non-linearities and element-wise additions.
\end{enumerate}

Each architecture's specification is modelled in Stream through the: a.) intra-core characteristics for each core (operand precision; PE array size; supported dataflow; memory hierarchy) and b.) inter-core characteristics (inter-core communication protocol: bandwidth-limited uni/bi-directional links going from one core's memory to another's or through a shared memory; inter-core connection topology; limited bandwidth off-chip link(s)). 

\textbf{Workload \& Performance Targets.} Each HW target has reported the measurements of their accelerator performance for different DNNs:
\begin{enumerate}
    \item The latency performance and memory footprint of FSRCNN~\cite{dong2016accelerating}, a fast super-resolution CNN, was measured on the single-core DepFIN processor.
    \item Jia et al. measured the latency performance of a ResNet50 segment.
    \item The latency performance and memory footprint of a ResNet18 segment was measured on the DIANA chip.
\end{enumerate}
Each measured DNN is modelled in Stream at the scheduling granularity supported by the hardware. The intra-core mapping and workload allocation is fixed in accordance with the reported measurements.

\textbf{Modeling Modes.}  
To demonstrate the importance of different modeling aspects, the validation is performed without and with communication modeling:
\begin{enumerate}
    \item A simple \emph{intra-core} model which takes into account the core mapping performances of all \revv{CTs}, which have potential temporal stalls because of insufficient bandwidth of the core memories.
    \item The detailed \emph{Stream} system model which uses \commname to model the core-to-core communication and off-chip accesses overhead. Furthermore, it facilitates detailed analysis of the memory footprint, offering valuable insights for expanded exploration of the design space.
\end{enumerate}

\begin{figure*}
\centering
\begin{subfigure}{.7\textwidth}
  \centering
  \includegraphics[width=\textwidth]{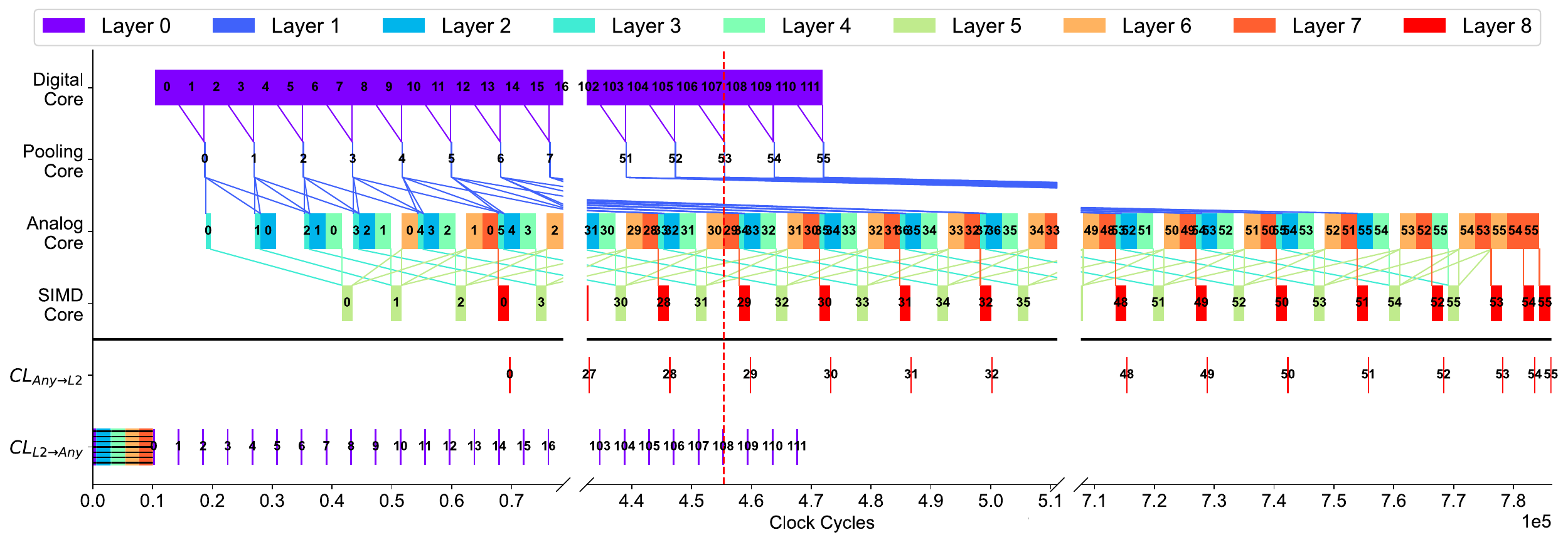}
  \caption{Schedule with fine-grain \revv{CT} dependencies and $L2$ communication}
  \label{fig:sub1}
\end{subfigure}%
\begin{subfigure}{.3\textwidth}
  \centering
  \includegraphics[width=\textwidth]{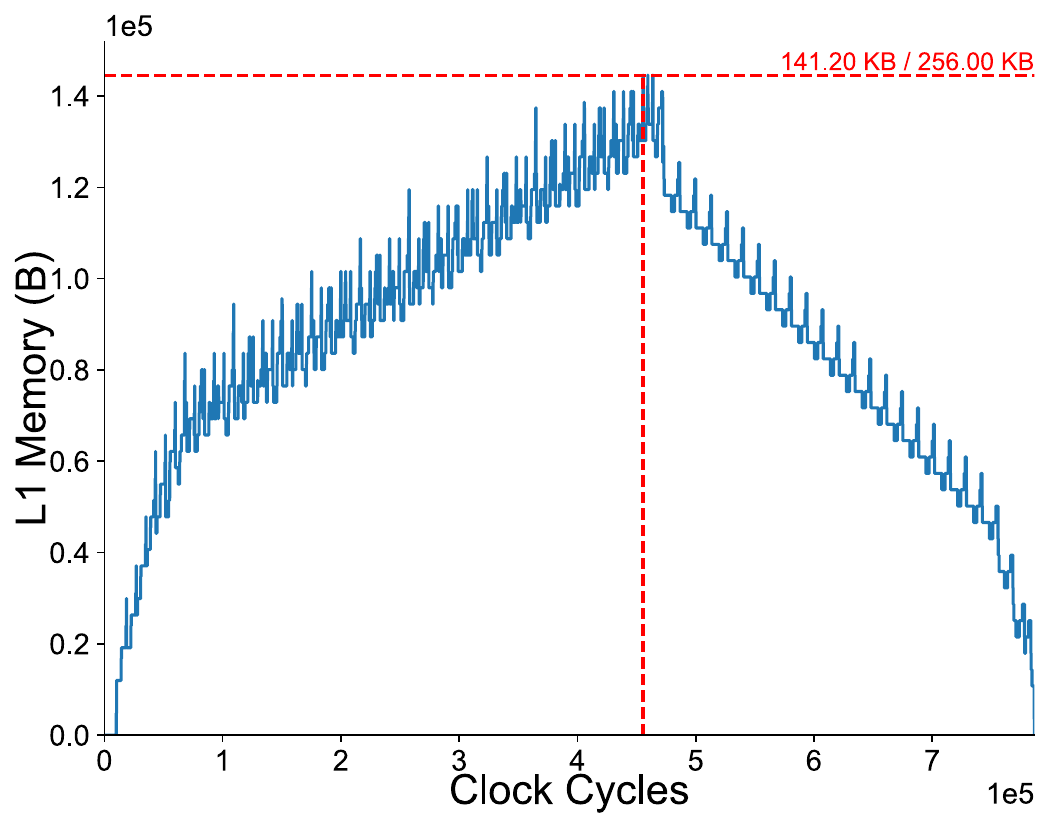}
  \caption{$L1$ memory usage}
  \label{fig:sub2}
\end{subfigure}
\caption{(a) Layer-fused schedule and (b) $L1$ memory usage of ResNet18 validation on DIANA.}
\label{fig:diana_schedule}
\vspace{-0.5em}
\end{figure*}

The outcomes of the validation process are presented in Table~\ref{tab:validation} and Fig.~\ref{fig:diana_schedule} visualizes the schedule and $L1$ memory usage of the DIANA results. Latency is characterized by the completion time of the final computation or communication task, measured in cycles. Concurrently, the memory footprint is determined by the peak memory utilization across the scheduling period, predicated on the assumption of 8-bit quantization for both activations and weights. Notably, Stream demonstrates superior precision in latency modeling these diverse architectures compared to the intra-core model, which neglects the overhead and contention for core-to-core and off-chip accesses.

\setlength\extrarowheight{2pt}
\begin{table}
\centering
\caption{Validation results for targeted architectures.}

    \begin{tabular}{|>{\centering}p{0.76in}||>{\centering}p{0.45in}||>{\centering}p{0.21in}|>{\centering}p{0.21in}||>{\centering}p{0.21in}|>{\centering}p{0.21in}|}
     \hline
     &\multicolumn{5}{|c|}{Latency Validation ($10^4$ clock cycles)}\tabularnewline
     \hline
     Architecture & Measured & \multicolumn{4}{|c|}{Modelled}\tabularnewline
     \hline
     & & \multicolumn{2}{|c||}{Intra-core} & \multicolumn{2}{|c|}{Stream} \tabularnewline
     \hline
     DepFiN \cite{depfin} & 618 & 502 & 81\% & 637 & 96\% \tabularnewline
     Jia et al. \cite{verma_jssc} & 6.6 & 6.2 & 93\% & 6.4 & 97\% \tabularnewline
     DIANA \cite{diana} & 81.2 & 78.1 & 95\% & 79.2 & 97\% \tabularnewline
     \hhline{|=||=||====|}
     &\multicolumn{5}{|c|}{Memory Validation (KB)}\tabularnewline
     \hline
     & & \multicolumn{2}{|c||}{Intra-core} & \multicolumn{2}{|c|}{Stream} \tabularnewline
     \hline
      DepFiN \cite{depfin} & 238 & \xmark & \xmark & 262 & 90\% \tabularnewline
     Jia et al. \cite{verma_jssc} & N/A & \xmark & \xmark & 36 & N/A \tabularnewline
     DIANA \cite{diana} & 134 & \xmark & \xmark & 114 & 85\% \tabularnewline
     \hline
    \end{tabular}
\label{tab:validation}
\vspace{-1em}
\end{table}


\section{Workload Allocation and Scheduling Optimization with \alloname}
\label{sec:allocation}

\begin{figure}[!t]
\centering
\includegraphics[width=\columnwidth]{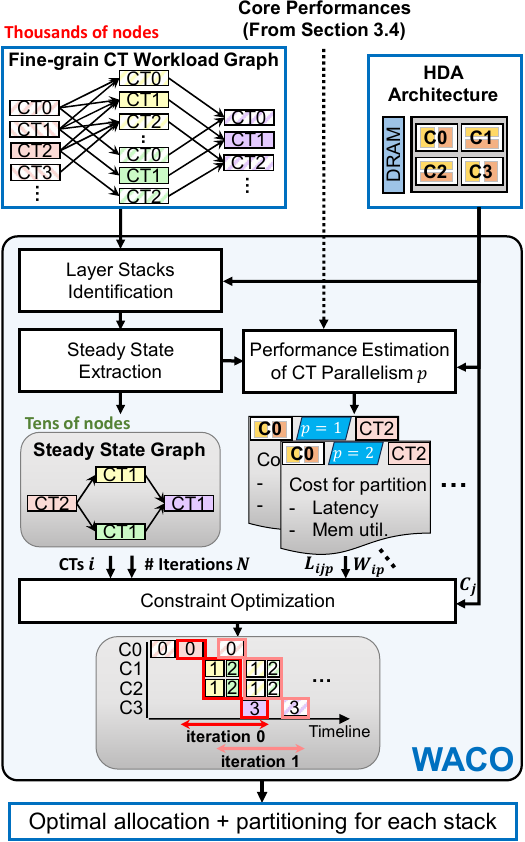}
\caption{The workload allocation is optimized using constraint optimization of the steady-state \revv{CTs} with the cost of different degrees of \revv{CT} parallelism in the $K$ dimension.}
\vspace{-1em}
\label{fig:waco}
\end{figure}

\subsection*{Goal}

The allocation and scheduling of partitioned workloads across available cores is a critical challenge for efficiently deploying DNNs on HDA architectures. Cores with different dataflows can be efficient for some DNN kernels but underutilized for others. Besides allocation, tiles must be scheduled correctly, accounting for fine-grain \revv{CT} dependencies. Fine-grain layer fusion can produce large workload graphs with up to $10^6$ tiles, making current solutions impractical without a more strategic approach.

\subsection*{Workload Allocation through Constraint Optimization}

To address this, Stream implements \alloname, a workload allocator and scheduler which uses constraint optimization to only allocate and schedule the \emph{steady-state} portion of the workload graph. 
This portion comprises the work within each layer that is repeated across numerous iterations of the layer-fused process. 
Fig.~\ref{fig:waco} highlights its building blocks. The workload graph is divided into one or more \emph{layer stacks} (Section~\ref{subsec:layer_stacks}) of which the steady state is extracted (Section~\ref{subsec:steady_state}). Furthermore, partitioning the \revv{CTs} across cores, referred to as \emph{\revv{CT} parallelism}, provides \alloname additional optimization capabilities (Section~\ref{subsec:k_partitioning}). A constraint optimization formulation accounts for potential overlap between steady-state iterations to efficiently interleave them (Section~\ref{subsec:co_formulation}).

\subsection{Layer Stacks Identification}
\label{subsec:layer_stacks}

The \revv{CT} workload graph is separated into multiple layer stacks. Within one layer stack, the execution of \revv{CTs} will be fused. In this work, a stack is limited to a contiguous set of layers whose combined weights don't exceed the HDA cores' memory capacity. This ensures weights of each layer can remain stationary on-chip throughout the layer-fused execution of the stack.

\subsection{Steady State Extraction}
\label{subsec:steady_state}

\begin{figure}[!t]
\centering
\includegraphics[width=\columnwidth]{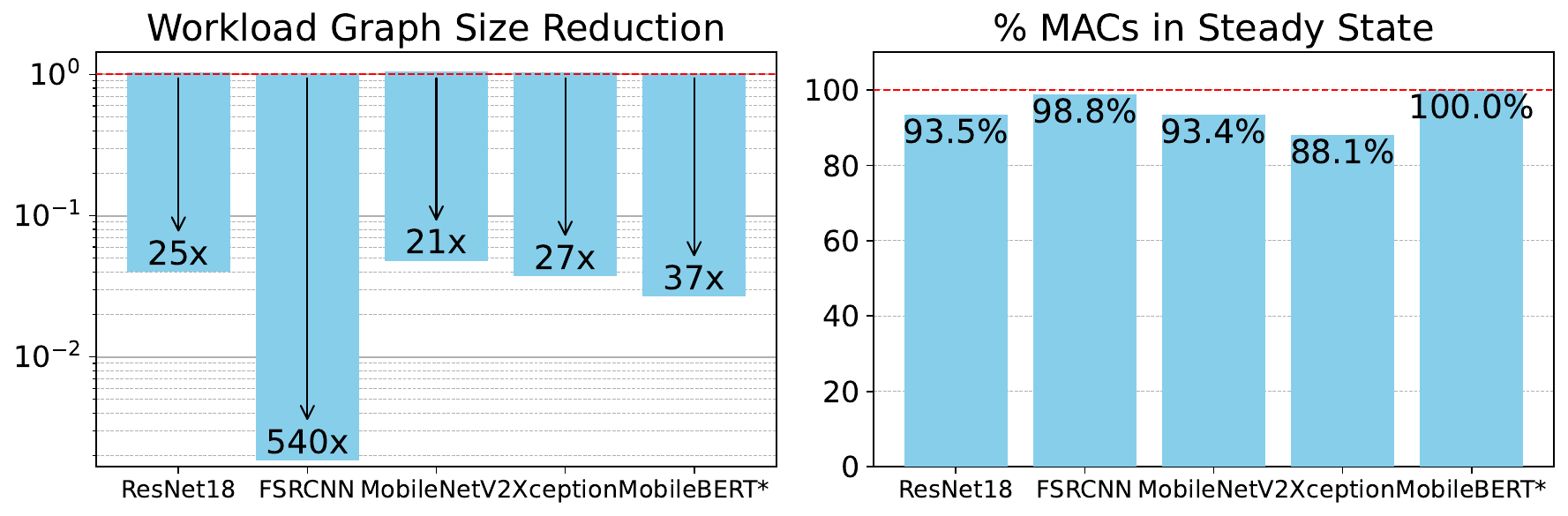}
\caption{The steady-state portion of different DNN workloads greatly reduces graph size while representing a large percentage of total workload MACs.} 
\vspace{-1em}
\label{fig:ss_percentages}
\end{figure}

For each stack, the steady-state portion is extracted by iterating through the \revv{CTs} of the last layer in the stack. For each iteration, the unique predecessors of the last layer's \revv{CTs} are retrieved and counted. These predecessors are the \revv{CTs} of each layer in the stack required to process the current sink \revv{CT} that have not been computed in a previous iteration. This is achieved using constant-time hash table lookups. If two sink \revv{CTs} have the same number of unique predecessors, they comprise the same work and their count is increased. The final counts are weighed by the number of MAC operations to find the iteration that is most significant. The \revv{CTs} of this iteration are called the \emph{steady state} and used for the subsequent steps. Fig.~\ref{fig:ss_percentages} demonstrates that considering only the steady state \revv{CTs} greatly reduces the size of the graph for line-based layer fusion, while still representing a large portion of the MAC operations of the entire workload. The greatest reduction is seen for FSRCNN, which can be separated into $540$ iterations of only $8$ \revv{CTs} thanks to its simple network structure and $540\times540$ input activation.

\subsection{Allocating \revv{Computation Tiles} to Multiple Cores}
\label{subsec:k_partitioning}

\alloname enables a \revv{CT} to be further parallelized across multiple cores in the output channel dimension $K$. This ensures that the weights —which are reused throughout the entire fused stack— remain distinct across cores. This parallelism factor, denoted with $p$, lies between 1 and the maximum number of cores the \revv{CT} can be allocated to and is required to be a divisor of $K$. To achieve an optimal allocation across cores while respecting constraints, it is necessary to calculate the weight requirement and estimate the performance of each partition for the possible $p$ values. The weight requirement for tile $i$ with parallelization factor $p$ is computed as: 

\begin{equation*}
W_{ip} = \frac{W_{i}}{p}
\end{equation*}

For latency, the \revv{CT} parallelization across cores can have an unexpected performance impact in HDAs, as \revv{1) the latency can be different on different cores}, and 2) some cores might require parallelization of $K$ within the core, thus causing array under-utilization when parallelizing across many cores.
\revv{These effects are automatically accounted for by \alloname: the latency $L_{ijp}$ of each CT $i$ on each core $j$ with parallelization factor $p$ is reduced by the same factor $p$ as long as the mapping onto the core leaves enough temporal loops $K_{ij}^{t}$ (the loops that remain after taking the within-core spatial parallelization into account). If there are not enough temporal loops, the latency reduction is limited. More formally: }


\begin{equation*}
L_{ijp} = 
\begin{cases} 
\frac{L_{ij}}{p} & \text{    if    } p \leq K_{ij}^{t} \\
\frac{L_{ij}}{K_{ij}^{t}} & \text{    if    } p > K_{ij}^{t}
\end{cases}
\end{equation*}

\rev{\subsection{Constraint Optimization}
\label{subsec:co_formulation}

\begin{figure}[!t]
\centering
\includegraphics[width=\columnwidth]{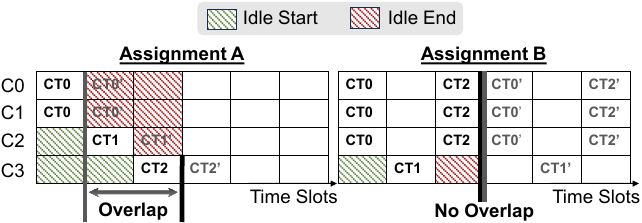}
\caption{Illustration of two different assignments of three steady-state tiles on four cores, leading to different overlap due to the idle start and end slots mismatch between cores.} 
\vspace{-1em}
\label{fig:waco_overlap}
\end{figure}

The core of \alloname's optimization is captured through a series of variables and constraints, formulated to delineate the scheduling and allocation of the steady-state \revv{CTs}. This problem is formulated and solved using Gurobi~\cite{gurobi} within the Stream framework.

\subsubsection{Objective Function}
The objective of the optimization is to minimize the total latency \(lat\) across \(N\) steady-state iterations, considering \((N-1)\) overlaps \(o\) between steady-state iterations, as illustrated in Fig.~\ref{fig:waco} and Fig.~\ref{fig:waco_overlap}.

\subsubsection{Constants}

The constants provided to the constraint formulation are:

\begin{itemize}
    \item \(W_{ip}\): The number of weights required for tile \(i\) given a \revv{CT} parallelism factor of \(p\).
    \item \(C_j\): The weight memory capacity of core \(j\).
    \item \(D_{i,i'}\): The dependencies between all tiles \(i\) and \(i'\).
    \item \(L_{ijp}\): The latency of tile \(i\) on core \(j\) given a \revv{CT} parallelism factor of \(p\).
\end{itemize}

\subsubsection{Variables}
The key base variables in the formulation are:

\begin{itemize}
    \item \(a_{ijs}\): Represents the binary assignment of tile \(i\) to core \(j\) during time slot \(s\). The latency of a time slot is not fixed and depends on the assignments. These form the basis of both the allocation and scheduling processes.
    \item \(p_{i}\): A one-hot binary vector of dimension $K_i$ indicating the chosen \revv{CT} parallelism for tile \(i\). It ensures that only one parallelism choice is selected for each tile.
    \item $IdleStart_{js}$ and $IdleEnd_{js}$: Binary variables that indicate for each timeslot $s$ and each core $j$ if the core is idle before any CT has been scheduled ($IdleStart$) or if the core is idle after all CTs have been scheduled ($IdleEnd$).
    \item Other variables are integer and derived from the base variables in the constraints below.
\end{itemize}

\subsubsection{Assignment, Dependency \& Weight Constraints}
The constraints are defined as follows:

\textbf{Assignment Constraints}
To ensure that no more than one tile is assigned to any core within a specific time slot, we impose the following constraint:
\begin{align}
    &\sum_{i} a_{ijs} \leq 1, &\forall j, \forall s \label{eq_a_1}
\end{align}

For the parallelism vector \(p_{i}\), we enforce that only one parallelism option is selected:
\begin{align}
    &\sum_{k} p_{ik} = 1 &\forall i \label{eq_k_1}
\end{align}

The total number of assignments across cores and slots for a tile \(i\) must match the selected parallelism, defined as:
\begin{align}
    &\sum_{j,s} a_{ijs} = \sum_{k} k \cdot p_{ik} &\forall i \label{eq_a_2}
\end{align}

\textbf{Slot and Dependency Constraints}
The slot \(Slot_{i}\) assigned to each tile \(i\) must respect the dependency constraints defined by the dependency matrix \(D\):
\begin{align}
    &Slot_{i} \leq Slot_{i'} - 1 &&\forall (i, i') \in D \label{eq_s_2}
\end{align}

\textbf{Weight Constraints}
The weights stored on each core~\(j\) are quantified and constrained to not exceed the weight memory capacity of each core:
\begin{align}
    &w_{j} = \sum_{i} \left(\sum_{k} \left(W_{ik} \cdot p_{ik}\right) \cdot \sum_{s} a_{ijs}\right) \leq C_{j} &\forall j \label{eq_w_1}
\end{align}

\subsubsection{Latency Constraints}
To accurately model the latency based on the assignments \(a\), we define the following variables:

\begin{itemize}
    \item \(y_{j}\): Represents the total number of assignments on each core:
    \begin{align}
        &y_{j} = \sum_{i,s} a_{ijs} &\forall j \label{eq_y_1}
    \end{align}
    \item \(z_{js}^{incl}\) and \(z_{js}^{excl}\): Represent the inclusive and exclusive sum of assignments on core \(j\) up to and including, or excluding, slot \(s\), respectively:
    \begin{align}
        &z_{js}^{incl} = \sum_{z=0}^{s} \sum_{i} a_{ijz}; \quad z_{js}^{excl} = \sum_{z=0}^{s-1} \sum_{i} a_{ijz} &\forall j, \forall s \label{eq_z_1}
    \end{align}
    \item \(l_{ij}\): Represents the latency of tile \(i\) on core \(j\), calculated by selecting the appropriate latency \(L_{ijk}\) based on the chosen parallelism:
    \begin{align}
        &l_{ij} = \sum_{s,k} a_{ijs} \cdot p_{ik} \cdot L_{ijk}, &\forall i, \forall j \label{eq_l_1}
    \end{align}
    \item \(l_{s}\): Represents the maximal latency for any tile assigned to slot \(s\) across all cores:
    \begin{align}
        &l_{s} = \max_{j} \left(\sum_{i} l_{ij} \cdot a_{ijs}\right) &\forall s \label{eq_l_2}
    \end{align}
    \item \(idle_{j}\): Represents the total start and end idle time on each core, as illustrated in Fig.~\ref{fig:waco_overlap}:
    \begin{align}
        &idle_{j} = \sum_{s} \left(IdleStart_{js} + IdleEnd_{js}\right) \cdot l_{s} &\forall j \label{eq_l_3}
    \end{align}
    \item \(o\): Represents the achievable overlap between steady-state iterations, defined as the minimum idle time across all cores:
    \begin{align}
        &o = \min_{j}(idle_{j}) \label{eq_l_4}
    \end{align}
    \item \(lat\): The total latency is computed as:
    \begin{align}
        &lat = N \cdot \sum_{s} l_{s} - (N - 1) \cdot o \label{eq_l_5}
    \end{align} 
\end{itemize}
}

\subsection{Comparison with SotA}

\begin{figure}[!t]
\centering
\includegraphics[width=\columnwidth]{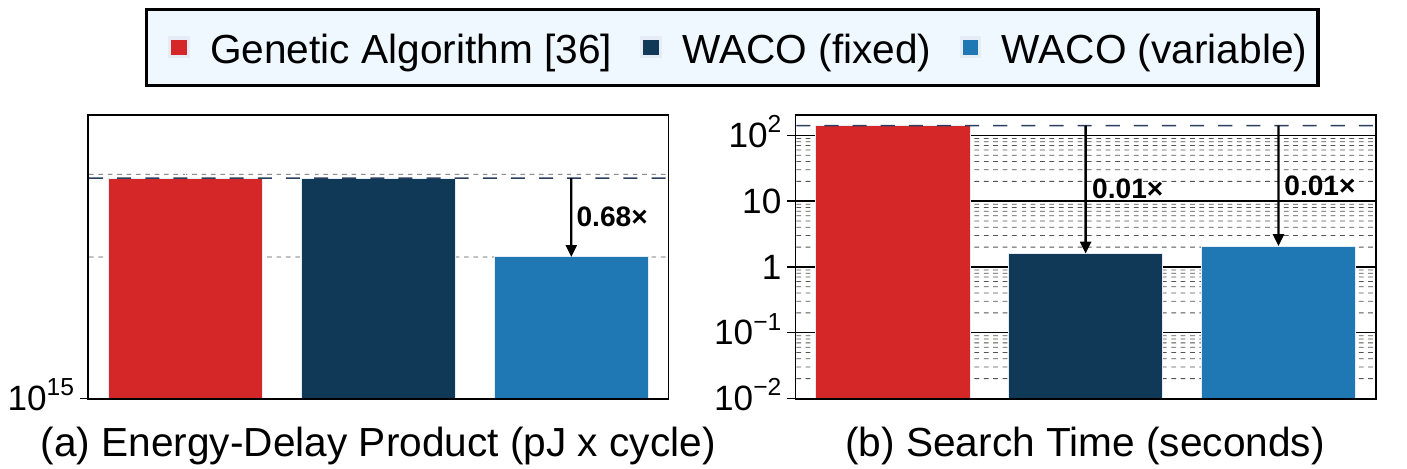}
\caption{Comparison of (a) energy-delay product and (b) required search time for the allocation and scheduling of ResNet18 on a quad-core architecture with four WS cores, using a state-of-the-art genetic algorithm-based optimizer and \alloname (see Table~\ref{tab:epxloration}).} 
\vspace{-1em}
\label{fig:resnet18_ga_vs_co}
\end{figure}

Fig.~\ref{fig:resnet18_ga_vs_co} provides a comparative analysis of the Energy-Delay Product (EDP) and search time for optimizing the allocation and scheduling of ResNet18 on a quad-core WS HDA. The analysis contrasts a state-of-the-art Genetic Algorithm (GA)-based search~\cite{karlgenetic} with two configurations of \alloname: one fixes $p=1$ to ensure a direct comparison with the GA approach, which does not support \revv{CT} parallelization across cores, and another allows variable \revv{CT} parallelism to explore the benefits of distributing a \revv{CT} across multiple cores to enhance utilization. \alloname with $p=1$ achieves similar EDP compared to the GA while significantly reducing the search time. Additionally, enabling \revv{CT} parallelization offers greater efficiency at only a marginal search time increase, reducing the EDP by $32 \%$ by filling idle slots present due to workload imbalances.
\section{Exploration}
\label{sec:exploration}

\begin{table*}[t]
\centering
\setlength\extrarowheight{2pt}
\caption{HDA architectures for the dataflow heterogeneity study. The architecture is varied from 1 to 8 cores with a total of $2^{12}$ PEs and \SI{4}{\mebi\byte} of on-chip memory.} 
\label{tab:epxloration}
\begin{tabular}{@{} cccccccccccccccccc @{}} 
 \toprule 
 \begin{tabular}{@{}c@{}}HDA \\ Arch.\end{tabular} & \begin{tabular}{@{}c@{}}No. \\ Arch.\end{tabular} & \multicolumn{9}{c}{\begin{tabular}{@{}c@{}}Weight Stationary Core Architecture \\ Fig.~\ref{fig:multi_core_architecture}(b)\end{tabular}} & \multicolumn{7}{c@{}}{\begin{tabular}{@{}c@{}}Output Stationary Core Architecture \\ Fig.~\ref{fig:multi_core_architecture}(c)\end{tabular}}\\
 \cmidrule(r){1-1} \cmidrule(lr){2-2} \cmidrule(lr){3-11} \cmidrule(l){12-18}
 & & PE Array & \multicolumn{4}{c}{Dataflow} & \multicolumn{2}{c}{Act Mem} & \multicolumn{2}{c}{W Mem} & PE Array & \multicolumn{2}{c}{Dataflow} & \multicolumn{2}{c}{Act Mem} & \multicolumn{2}{c}{W Mem} \\
 \cmidrule(lr){3-3} \cmidrule(lr){4-7} \cmidrule(lr){8-9} \cmidrule(lr){10-11} \cmidrule(lr){12-12} \cmidrule(lr){13-14} \cmidrule(lr){15-16} \cmidrule(l){17-18}
 & & row $\times$ col & C & FX & FY & K & Size & BW & Size & BW & row $\times$ col & OX & K & Size & BW & Size & BW \\
 & & & & & & & \SI{}{\mebi\byte} & \SI{}{\byte} & \SI{}{\mebi\byte} & \SI{}{\byte} & & & & \SI{}{\mebi\byte} & \SI{}{\byte} & \SI{}{\mebi\byte} & \SI{}{\byte} \\
 \midrule 
 1 Core  & 2 & 72 $\times$ 64 & 8 & 3 & 3 & 64 & 2 & 72 & 2 & 128 & 64 $\times$ 64 & 64 & 64 & 2 & 64 & 2 & 64\\
 2 Core  & 3 & 36 $\times$ 64 & 4 & 3 & 3 & 64 & 1 & 36 & 1 & 128 & 32 $\times$ 64 & 32 & 64 & 1 & 32 & 1 & 64 \\
 4 Core  & 5 & 36 $\times$ 32 & 4 & 3 & 3 & 32 & 0.5 & 36 & 0.5 & 64 & 32 $\times$ 32 & 32 & 32 & 0.5 & 32 & 0.5 & 32 \\
 8 Core & 9 & 18 $\times$ 32 & 2 & 3 & 3 & 32 & 0.25 & 18 & 0.25 & 64 & 16 $\times$ 32 & 16 & 32 & 0.25 & 16 & 0.25 & 32 \\
 \bottomrule 
\end{tabular}
\vspace{-1em}
\end{table*}

In this section, we deploy Stream to co-explore the optimal allocation in combination with architectural decisions across different workloads in two exploration studies. The studies are repeated for five DNN workloads: ResNet18~\cite{resnet}, FSRCNN~\cite{dong2016accelerating}, MobileNetV2~\cite{sandler2018mobilenetv2}, Xception~\cite{chollet2017xception} \rev{and the first body of MobileBERT~\cite{sun2020mobilebert}. 
The framework is evaluated offline, with parallel processing of different input combinations: workload, HDA and scheduling paradigm.}

The HDA architectures used in the different studies are built from two core architectures: a weight and an output stationary accelerator core. The core architectures are shown in Fig.~\ref{fig:multi_core_architecture}, with the specific PE array shape, dataflow, memory capacities and memory bandwidths specified in Table~\ref{tab:epxloration}. All memory read costs and write costs are extracted  in \SI{22}{\nano\meter} using CACTI~7~\cite{balasubramonian2017cacti}, an open-source memory profiling tool. Each PE includes a \SI{4}{\byte} register file. The output register file of each WS column is \SI{128}{\byte}. The output local buffer of the OS core is matched in capacity (\SI{128}{\byte} times the number of columns) and bandwidth (read-out of one row at \SI{16}{\bit} precision as opposed to writing of two rows at \SI{8}{\bit} in WS). 

The cores are interconnected through a core-to-core communication bus of $256$ bits per clock cycle. A single port of $128$ bits per clock cycle models contention of off-chip accesses to a \SI{256}{\mebi\byte} DRAM memory.

For all experiments, \alloname optimizes the allocation with variable \revv{CT} parallelism, which is then evaluated by \commname to obtain latency and energy efficiency estimations.  

\subsection{Workload Diversity \& Scheduling Granularity}
\label{subsec:cs1}

First, the benefits of layer fusion compared to layer-by-layer scheduling are analyzed for the five workloads. 

\begin{figure}[!t]
\centering
\includegraphics[width=\columnwidth]{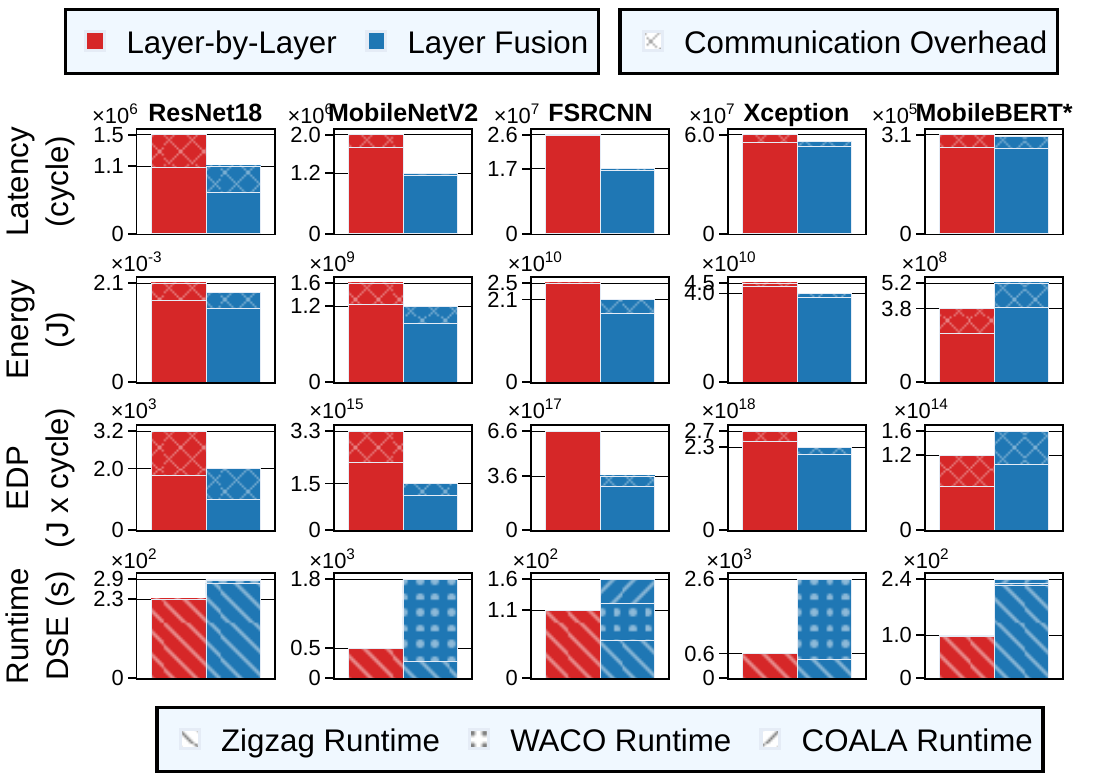}
\caption{Layer-by-layer vs. layer fusion performance of \alloname-optimized allocation for a quad-core HDA with four WS cores as in Table~\ref{tab:epxloration}. \footnotesize{* Exploration of first transformer block}} 
\vspace{-1em}
\label{fig:exploration_study_1}
\end{figure}

Fig.~\ref{fig:exploration_study_1} compares the latency, energy, EDP and DSE runtime of the best found allocation by \alloname onto a quad-core HDA with four WS cores. The highlighted sections denote the communication overheads. MobileNetV2 and FSRCNN benefit largely from layer fusion, as they are activation-dominant. Their EDP is reduced by $2.2\times$ and $1.8\times$, respectively, under layer fusion, thanks to the small weight size which allows \alloname to fuse many layers into a single stack, bringing large benefits as all intermediate activation can be stored on-chip. MobileBERT challenges the current fusion strategy due to the complexity of its attention mechanism and reduced energy efficiency during layer fusion, caused by limited weight reuse in dense matrix multiplication layers. Stream offers potential for exploring new fusion strategies and specialized dataflows to address these issues, making it a viable option for future research.

\subsection{Dataflow Heterogeneity}


\begin{figure*}[h]
\centering
\includegraphics[width=\textwidth]{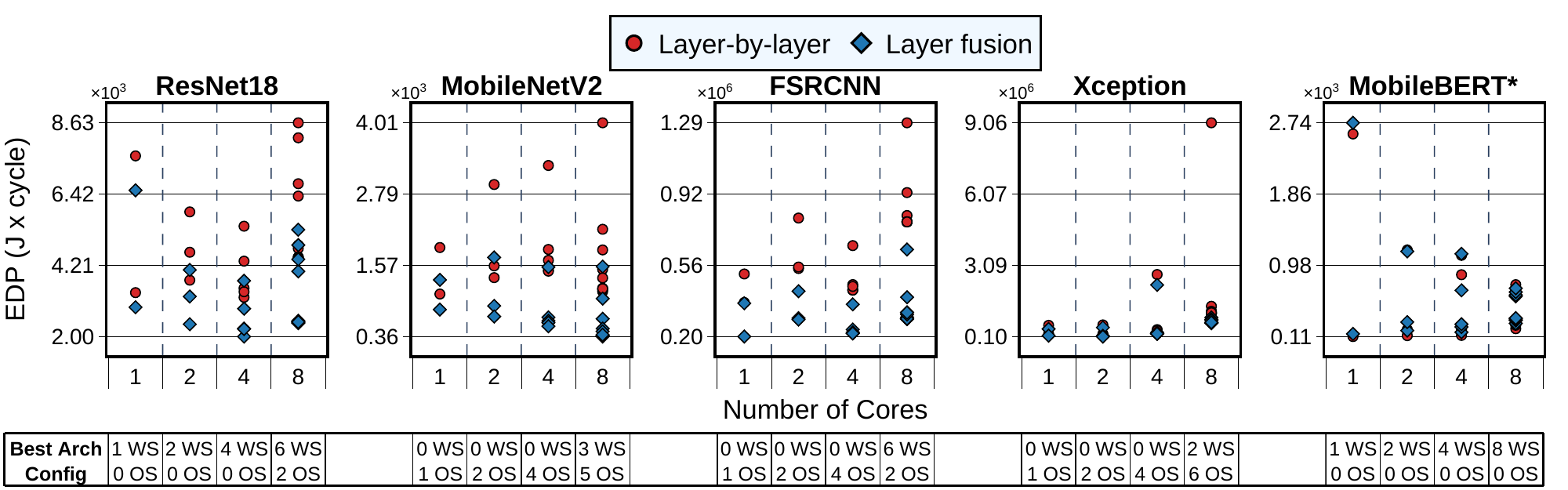}
\caption{\rev{Exploration results showing EDP impact of varying scheduling, number of cores and dataflow heterogeneity within a fixed area of $2^{12}$ PEs and \SI{4}{\mebi\byte}.}} 
\vspace{-1em}
\label{fig:exploration_study_2}
\end{figure*}

The initial exploration maintained a uniform dataflow across all four cores, resulting in a homogeneous architecture. This demonstrated that layer fusion is advantageous for various workloads; however, the system's performance also significantly depends on the number of cores and the specific dataflow employed in each core. Consequently, a subsequent study evaluates the performance implications of varying both the number of cores and their dataflows. This study explores a range of architectures from a single core up to eight cores, all constrained to the same total number of processing elements (PEs) and memory capacity, thereby maintaining an iso-area condition. For each core count, the study examines $\binom{n}{k}$ possible combinations of architecture configurations, where $n=2$ represents the two types of cores available, and $k$ is the number of cores in the system.

Fig.~\ref{fig:exploration_study_2} presents a comparison of Energy-Delay Product (EDP) performances for different workloads under layer-by-layer and layer fusion strategies across various core counts and dataflow configurations. The optimal architecture for each scenario is shown below, revealing that distinct workloads exhibit preferences for specific dataflows. Notably, MobileNetV2, characterized by numerous small layers, shows improved performance with a greater number of OS cores, which are more effective for depthwise convolutional layers. Conversely, other networks experience diminished performance in larger core counts due to the increased overhead associated with core-to-core communication.

\begin{figure}[h]
\centering
\includegraphics[width=\columnwidth]{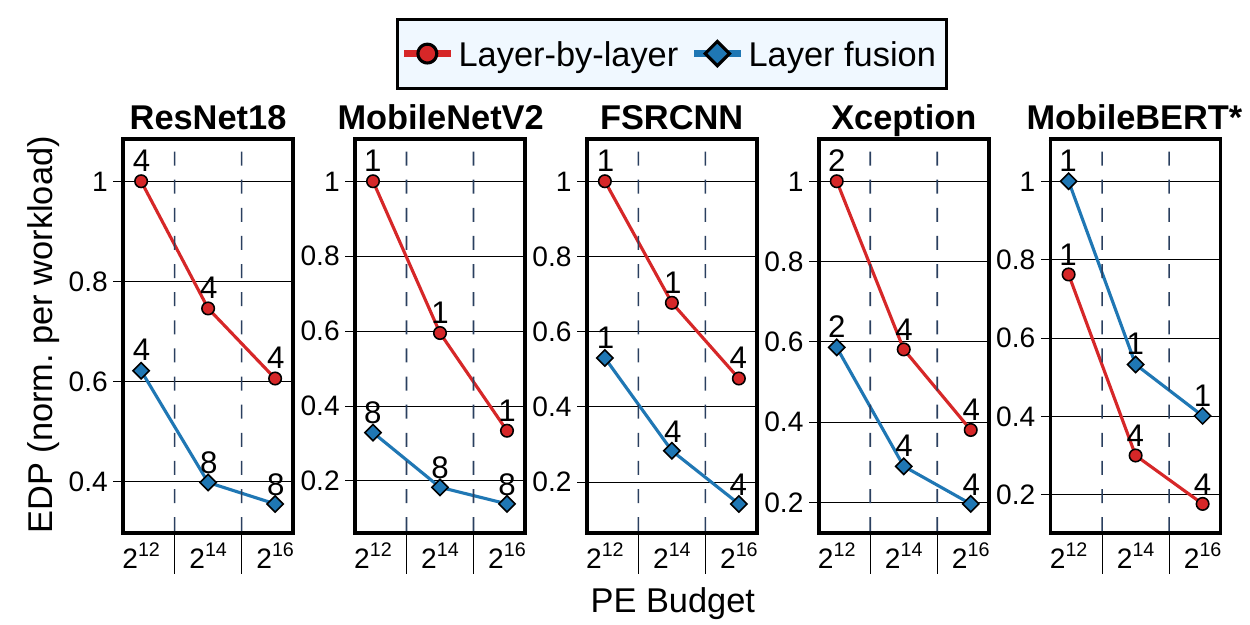}
\caption{\rev{EDP performance improvements of the hardware scale-up exploration. The annotations show increasing optimal core counts for larger budgets.}} 
\vspace{-1em}
\label{fig:exploration_study_3}
\end{figure}

\rev{\subsection{Hardware Scale-up}
Finally, the hardware is scaled to larger sizes to evaluate the interaction between the available compute elements and the optimal number of cores by increasing the PE budget from $2^{12}$ to $2^{14}$ and $2^{16}$. The dataflows are scaled similarly to Table~\ref{tab:epxloration}, keeping the array as square as possible. While for smaller PE budgets, a single-core system may achieve the best utilization, this can change as the array size increases, particularly if the layers do not exhibit sufficient parallelism to fully leverage the expanded array anymore.

Fig.~\ref{fig:exploration_study_3} summarizes the results of the scaled out simulations across varying compute budgets. For each PE budget, the EDP-optimal architecture configuration —comprising core counts and dataflow per core— is presented across various workloads and scheduling paradigms. The annotated number represents the optimal core count. The observed increase in optimal core counts as PE budgets expand highlights the strong relationship between architecture configuration, workload type, and compute resources.

}

\newpage
\section{Conclusion}

This work presented \emph{Stream}, a design space framework for the optimization of layer-fused deep neural networks on heterogeneous dataflow accelerators, addressing the challenges posed by ever-growing network sizes and the need for reduced latency and energy consumption at the edge. 

Central to Stream's implementation is a communication and memory-aware latency analysis (\commname), which enables rapid performance estimation in terms of latency and energy efficiency. \commname is validated against three state-of-the-art accelerator systems that employ layer fusion. A workload allocation engine (\alloname) strategically allocates computational cores to maximize efficiency. \alloname leverages constraint optimization techniques that, in combination with the parallelism and memory efficiency afforded by layer fusion, significantly enhance hardware utilization. This demonstrated a reduction of up to $2.2\times$ in energy-delay product compared to traditional layer-by-layer scheduling.


Moreover, the flexibility of Stream offers a comprehensive tool for future research and development in efficient DNN deployment on evolving accelerator architectures.

\ifCLASSOPTIONcompsoc
  \section*{Acknowledgments}
\else
  \section*{Acknowledgment}
\fi
The authors thank Koen Goetschalckx for his valuable comments. This project has been partly funded by the European Research Council (ERC) under grant agreement No.~101088865, the European Union’s Horizon 2020 programme under grant agreement No.~101070374, the Flanders AI Research Program and KU Leuven.

\ifCLASSOPTIONcaptionsoff
  \newpage
\fi



%


\bibliographystyle{IEEEtran}
\bibliography{refs}

%

\begin{IEEEbiography}
    [{\includegraphics[width=1in,height=1.25in,clip,keepaspectratio]{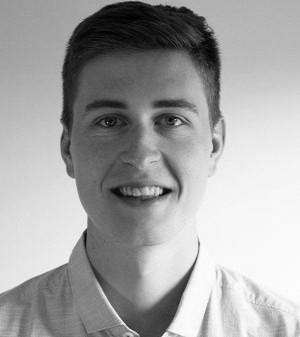}}]{Arne Symons}
received the BSc and MSc degree in electrical engineering from KU Leuven, Leuven, Belgium in 2018 and 2020, respectively. He is currently working towards a PhD degree in mapping optimization of machine learning workloads on specialized accelerator architectures. In 2022 he was a visiting researcher at the Robust Systems Group of Stanford University, and interned as a research scientist at Meta in 2023.
\end{IEEEbiography}

\begin{IEEEbiography}
    [{\includegraphics[width=1in,height=1.25in,clip,keepaspectratio]{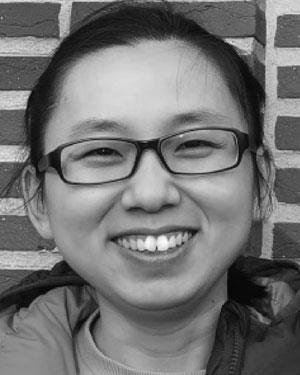}}]{Linyan Mei}
received the BSc degree in electronic design and technology from the Beijing Institute of Technology (BIT), Beijing, China, in 2016, and the MSc and PhD degree in electrical engineering from KU Leuven, Leuven, Belgium, in 2018 and 2023, respectively. Her PhD thesis is entitled \textit{Design Space Exploration of Deep Learning Accelerators}. She was an intern with imec, Leuven, Belgium, from 2017 to 2018 and with Meta in 2021. She is currently working at Huawei in China on hardware-software co-design of machine learning workloads.
\end{IEEEbiography}

\begin{IEEEbiography}
    [{\includegraphics[width=1in,height=1.25in,clip,keepaspectratio]{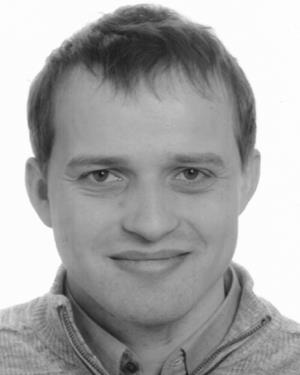}}]{Steven Colleman}
received the BSc, MSc and PhD degree in electrical engineering from KU Leuven, Leuven, Belgium, in 2016, 2018 and 2024 respectively. His PhD thesis is entitled \textit{Modeling and analysis for efficient hardware mapping of neural network algorithms}. He is currently working at Axelera AI in Leuven, Belgium.
\end{IEEEbiography}

\begin{IEEEbiography}
    [{\includegraphics[width=1in,height=1.25in,clip,keepaspectratio]{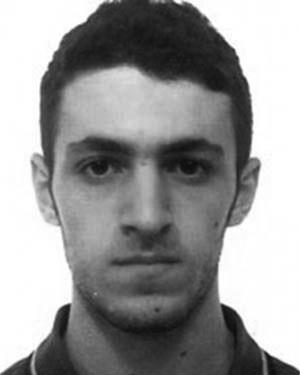}}]{Pouya Houshmand}
received the BSc and MSc degrees in electrical engineering from the Polytechnic of Turin, Italy, in 2017 and 2019, respectively. He is currently working towards a PhD degree in architectures for deep neural network accelerators in the ESAT-MICAS Laboratories, KU Leuven, Leuven, Belgium. His current research interests include algorithm-hardware co-design, in-memory computing, and emerging technologies.
\end{IEEEbiography}

\begin{IEEEbiography}
    [{\includegraphics[width=1in,height=1.25in,clip,keepaspectratio]{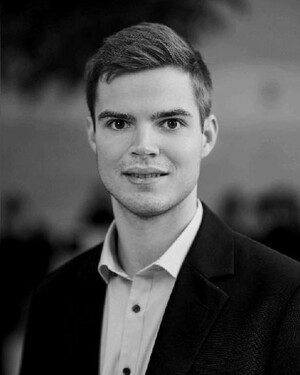}}]{Sebastian Karl}
received the BSc and MSc degrees in electrical engineering from the Technical University of Munich, Germany, in 2020 and 2022, respectively. He received the MSc Management degree from Lund University School of Economics and Management in 2023. He is currently working as a project manager and business developer for the business field Neuromorphic Computing at Fraunhofer IIS in Erlangen, Germany.
\end{IEEEbiography}

\begin{IEEEbiography}
    [{\includegraphics[width=1in,height=1.25in,clip,keepaspectratio]{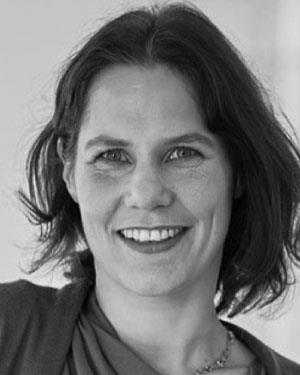}}]{Marian Verhelst}
(\IEEEmembership{Fellow,~IEEE}) received the PhD degree from KU Leuven, Leuven, Belgium, in 2008. She is a full-time professor with ESAT-MICAS Laboratories, KU Leuven. Her research focuses on embedded machine learning, hardware accelerators, and low-power edge processing. Before that, she worked with Intel Labs, US, from 2008 until 2011. She is an SSCS distinguished lecturer, was a member of the Young Academy of Belgium, an associate editor of the \textit{IEEE Transactions on Very Large Scale Integration (VLSI) Systems}, \textit{IEEE Transactions on Circuits and Systems II: Express Briefs}, and \textit{IEEE Journal of Solid-State Circuits}. She currently holds an ERC grant from the European Union and was the laureate of the Royal Academy of Belgium.
\end{IEEEbiography}







\end{document}